\def\ra{\rightarrow}
\def\Ra{\Rightarrow}

\def\({\left(}
\def\){\right)}
\def\[{\left[}
\def\]{\right]}

\def\a{\alpha}
\def\be{\beta}

\def\G{\Gamma}
\def\de{\delta}

\def\la{\lambda}

\def\th{\theta}
\def\om{\omega}
\def\Om{\Omega}
\def\vf{\varphi}

\def\S{\Sigma}
\def\z{\zeta}

\newcommand\bm[1]{\mbox{\boldmath$#1$}}

\def\build#1_#2^#3{\mathrel{\mathop{\kern
																			0pt#1}\limits_{#2}^{#3}}}

\def\dfrac#1#2{{\displaystyle{#1\over#2}}}

\def\sqr#1#2{{\vcenter{\vbox{\hrule height.#2pt
									\hbox{\vrule width.#2pt height#1pt \kern#1pt
            \vrule width.#2pt}
         \hrule height.#2pt}}}}

\def\coma{\quad ,\quad}
\def\dps{\displaystyle}

\def\beq{\begin{equation}}

\def\eeq{\end{equation}}

\def\beqa{\begin{eqnarray}}

\def\eeqa{\end{eqnarray}}

\def\ba{\begin{array}}

\def\ea{\end{array}}

\def\beqas{\begin{eqnarray*}}

\def\eeqas{\end{eqnarray*}}

\documentclass[10pt,a4paper,twoside]{article}

\usepackage{indentfirst}			

\usepackage{graphicx}				

\usepackage{amsgen,amsfonts,amssymb,amsbsy}	

\newenvironment{resum}{\begin{quote}\small}{\end{quote}}

\newcommand{\bfsf}[1]{\textsf{\textbf{#1}}}

\begin{document}

\thispagestyle{plain}		

\begin{center}


{\LARGE\bfsf{An approximate global stationary \\
axisymmetric solution of Einstein equations}}

\bigskip


\textbf{J. Mart\'\i{}n--Mart\'\i{}n}$^1$,
\textbf{A. Molina}$^2$ and \textbf{E. Ruiz}$^1$


$^1$\textsl{Universidad de Salamanca, Spain.} \\
$^2$\textsl{Universitat de Barcelona, Spain.}

\end{center}

\medskip


\begin{resum}
We present an approximate global solution of Einstein's  equations which des\-cribes
a simple star model:  a self-gravitating perfect fluid with constant density and rigid
rotation (Born). In order to do this we carry out a second order post--minkowskian
approximation of the problem, and then we use a slow rotation subapproximation. The
resulting metric depends on three arbitrary constants: density, rotational velocity
and stellar radius in the non-rotating limit. Mass, angular momentum, quadrupole and
other moments are expressed in terms of these three constants. As a side result we
show that this type of fluid cannot be a source of the Kerr metric.
\end{resum}

\bigskip




\section{Introduction}

In this work we present an approximate solution of the Einstein equations descri\-bing a
global model for the gravitational field generated by a bounded, self--gravitating
stationary and axisymmetric body rotating rigidly with constant angular velocity.
The exterior metric corresponds to an asymptotically flat stationary, axisymmetric
vacuum solution, while the interior is a perfect--fluid solution with the  above
pro\-perties. Both solutions (exterior and interior) are matched at the surface of
zero pressure using the Lichnerowicz matching conditions, i. e., the global solution
is $C^1$ at the boundary. To carry out the matching we use global  harmonic coordinates,
since these coordinates provide the necessary arbitrary constants.
\subsection{Generic expresions of the metric}

Let us consider a stationary me\-tric with axial symmetry. As usual,
the me\-tric tensor  will be  referred to a system of spherical
coordinates $\{t,r,\th,\vf\}$ adapted to the symmetries and to the
Papapetrou structure, so that the radial and polar coordinates
are defined  on the 2-D surfaces orthogonal to the  surfaces of transitivity. For
convenience, we shall use an  orthonormal cobasis, with respect to the Minkowskian
metric, associated to these coordinates:

\beqa
{\bf g} &\!\!\!\!=& \!\!\!\!g_{00}\,{\bm\om^0}{\otimes\,}{\bm\om^0} +
g_{03}\({\bm\om^0}{\otimes\,}{\bm\om^3} +{\bm\om^3}{\otimes\,}{\bm\om^0}\)+
g_{33}\,{\bm\om^3}{\otimes\,}{\bm\om^3} \nonumber\\
 &\ \ +&\!\!\!\!g_{11}\,{\bm\om^1}{\otimes\,}{\bm\om^1} +
g_{12}\({\bm\om^1}{\otimes\,}{\bm\om^2}+{\bm\om^2}{\otimes\,}{\bm\om^1}\) +
 g_{22}\,{\bm\om^2}{\otimes\,}{\bm\om^2}
\eeqa

\beq
{\bm\om^0}=dt \coma {\bm\om^1}=dr \coma {\bm\om^2}=r\, d\th \coma {\bm\om^3} =
r\sin\th\, d\vf
\label{cobase}
\eeq
However, for computational reasons, we shall write the metric  tensor ${\bf g}$ with
respect to another tensor basis

\beq
{\bf g}\, = g_{00}{\bf t^{00}} + g_{03}{\bf t^{03}} + f_d {\bf t^d} + f_0
{\bf t^0} + f_1 {\bf t^1} +f_2 {\bf t^2}
\eeq

\beq
\left\{\ba{ll}
{\bf t^{00}} &\!\!\!\equiv {\bm\om^0}{\otimes\,}{\bm\om^0} \\
\vspace*{-3.5mm}\\
{\bf t^{03}} &\!\!\!\equiv{\bm\om^0}{\otimes\,}{\bm\om^3} + 
{\bm\om^3}{\otimes\,}{\bm\om^0}\\
\ea\right.
\quad
\left\{\ba{ll}
{\bf t^d} &\!\!\!\equiv \delta_{ij}\,dx^i{\otimes\,} dx^j \\
\vspace*{-3.5mm}\\
{\bf t^0} &\!\!\!\equiv (\delta_{ij}-3e_i e_j)dx^i{\otimes\,} dx^j \\
\vspace*{-3.5mm}\\
{\bf t^1} &\!\!\!\equiv (k_ie_j + k_je_i)dx^i{\otimes\,} dx^j \\
\vspace*{-3.5mm}\\
{\bf t^2} &\!\!\!\equiv (k_ik_j -m_jm_i)dx^i{\otimes\,} dx^j \\
\ea\right.
\label{basetensor}
\eeq
where $\{x^i\}$ are the ``cartesian" coordinates associated to the spherical ones
$\{r,\th,\vf\}$ and the covectors $\{ k_i, e_i, m_i\}$ are defined as

\beq
d(r\sin\th)= k_i\, dx^i \coma
dx^3 = e_i\,dx^i \coma
\om^3 = r\sin\th\,d\vf = m_i\,dx^i
\eeq
that is to say, they represent respectively the unitary radial, axial and azimuthal
cylindrical covectors. The tensor basis $\{\bf t^{00},t^{03},t^d,\bf t^0, t^1,
t^2\}$ appears quite naturally in the linear  and higher order approximations of
Einstein's equations, which require a Poisson-like equation to be solved, which in
turn requires the introduction of the tensor spherical harmonics (we shall see that
they are directly related to the tensor basis we are considering). Notice that the
tensor ${\bf t^d}$ is pure trace, while $\{\bf t^0, t^1, t^2\}$ are traceless (with
respect to the flat 3--dimensional metric).

The relationship between the  metric components on  the spherical co-basis
(\ref{cobase}) and the coefficients of the metric with respect to the ``pure
trace--traceless" tensor basis (\ref{basetensor}) is:

\beq\ba{ll}
f_d \!=& \!\!\!\dfrac13(g_{11}+g_{22}+g_{33}) \\[2mm]
f_0 \!=& \!\!\!\dfrac16\[g_{11}(1-3\cos^2\theta)+3g_{12}\sin 2\theta
+g_{22}(1-3\sin^2\theta)+ g_{33}\] \\[2mm]
f_1 \!=& \!\!\!\dfrac12(g_{11}-g_{22})\sin 2\theta + g_{12}\cos 2\theta \\[2mm]
f_2 \!=& \!\!\!\dfrac12\(g_{11}\sin^2\theta+g_{12}\sin 2\theta
+g_{22}\cos^2\theta -g_{33}\) \\
\ea\eeq
\subsection{Post--minkowskian approximation}

In this section we shall briefly describe  the post--minkowskian
approximation using the harmonic condition (we could also use the so called
q--armonic condition \cite{abmmr}).  We begin by defining the deviation of the metric
in ``cartesian" coordinates

\beq
h_{\a\be} := g_{\a\be} - \eta_{\a\be} 
\eeq
Let us consider the Einstein equations and harmonicity conditions

\beq
\left\{\!\!\!\!\!\!\!\ba{ll}
&R_{\a\be} = 8\pi\(T_{\a\be}-\frac12 g_{\a\be}T\) \equiv 8\pi {\cal T}_{\a\be}\\
\vspace*{-1mm}\\
&g^{\la\mu}\G^\rho_{\la\mu} = 0 \\
\ea\right.
\eeq
(we assume $G=c=1$). As it is known the Ricci tensor and the l.h.s. of the harmonic
condition can be written in terms of the deviation $h_{\a\be}$ as follows:

\beq
R_{\a\be} = -\frac12\square\, h_{\a\be} +
\frac12\partial_\a\!\[\partial^\rho\!\(h_{\rho\be}-\frac12\eta_{\rho\be}\,h\)\] 
+ \frac12\partial_\be\!\[\partial^\rho\!\(h_{\rho\a}-\frac12\eta_{\rho\a}\,h\)\] 
+N_{\a\be}
\label{Ricci}
\eeq

\beq
\G_\a \equiv \eta_{\a\rho}g^{\la\mu}\G^\rho_{\la\mu} =
\partial^\rho\!\(h_{\rho\a}-\frac12\eta_{\rho\a}\,h\) + H_\a
\label{harmo}
\eeq
where $N_{\a\be}$ and $H_\a$ are at least quadratic  in $h_{\a\be}$. We
 also use the notation $h\equiv \eta^{\a\be}h_{\a\be}\,$, $\partial^\rho \equiv
\eta^{\rho\la}\partial_\la$ and $\,\square\equiv \eta^{\la\mu}\partial_{\la\mu}$
(usual d'Alembertien operator).  Consequently the Einstein equations using harmonic 
coordinates can be written as:

\beq
\left\{\!\!\!\!\!\!\!\ba{ll}
& \square\,h_{\a\be} = -16\pi {\cal T}_{\a\be} + 2\[N_{\a\be} - \partial_{(\a}H_{\be)}\]
\\
\vspace*{-1mm}\\
& \partial^\rho\!\(h_{\rho\a}-\frac12\eta_{\rho\a}\,h\) = -H_\a \\
\ea\right.
\eeq
and then the equations for the {\em linear stationary approximation\/} can be written as
follows (where $\triangle$ is the flat laplacian operator):

\beq
\left\{\!\!\!\!\!\!\!\ba{ll}
&\triangle h^{(1)}_{\a\be} = -16\pi {\cal T}^{(1)}_{\a\be} \\
\vspace*{-2mm}\\
&\partial^k \[h^{(1)}_{k\a} -\frac12 h^{(1)}\eta_{k\a}\,\] = 0 \\
\ea\right.
\label{lineareq}
\eeq
where, as shown below, the tensor ${\cal T}^{(1)}_{\a\be}$ associated to the
energy--momentum tensor requires only the flat Minkowski metric. Once  a solution
of these equations is fixed we can solve the {\em postlinear approximation\/} from the
equations

\beq
\left\{\!\!\!\!\!\!\!\ba{ll}
&\triangle h^{(2)}_{\a\be} = -16\pi {\cal T}^{(2)}_{\a\be}+2N^{(2)}_{\a\be}  -
\partial_\a H^{(2)}_\be -
\partial_\be H^{(2)}_\a \\
\vspace*{-2mm}\\
&\partial^k \[h^{(2)}_{k\a} -\frac12 h^{(2)}\,\eta_{k\a}\] = -H^{(2)}_\a \\
\ea\right.
\label{plineareq}
\eeq
where ${\cal T}^{(2)}_{\a\be}\,$, $\,N^{(2)}_{\a\be}\,$ and $\,H^{(2)}_\a$ are
constructed from the linear deviation $h^{(1)}_{\a\be}$. Note that using the  expressions
(\ref{Ricci},\ref{harmo}) it obviously follows  that  $\,N^{(2)}_{\a\be}\,$ and
$\,H^{(2)}_\a$ may be obtained in an algorithmic way easily implemented on a computer.
One only needs evaluate this expressions for the metric $g^{[1]}_{\a\be}\equiv
\eta_{\a\be}+h^{(1)}_{\a\be}$

\section{Exterior solution}
\subsection{Linear approximation (exterior solution)}

The general asymptotically flat solution of   equations (\ref{lineareq}) in vacuum
(exterior) with the assumed symmetries is

\beqa
{\bf h}^{(1)}_{\rm ext} &\!\!\!=& \!\!\!2\sum_{n=0}^\infty
\frac{M_n^{(1)}}{r^{n+1}}{\bf T}_n +  2\sum_{n=1}^\infty \,\frac{J_n^{(1)}}{r^{n+1}}{\bf
Z}_n  + 2\sum_{n=0}^\infty \frac{M_n^{(1)}}{r^{n+1}}{\bf D}_n \nonumber\\
&+&\!\!\!\sum_{n=1}^\infty\frac{A_n^{(1)}}{r^{n+1}}{\bf E}_n 
+\sum_{n=2}^\infty\frac{B_n^{(1)}}{r^{n+1}}{\bf F}_n 
\label{linearextsol}
\eeqa
where the notation:

\beq
\left\{\ba{ll}
{\bf E}_n &\!\!\equiv \,-\dfrac23n{\bf D}_n +\dfrac23n{\bf H}^0_n\, +
{\bf H}^1_n \\
\vspace*{-3.5mm}\\
{\bf F}_n &\!\!\equiv \,\dfrac12n(n-1){\bf H}^0_n 
+(n-1){\bf H}^1_n -
\dfrac12 {\bf H}^2_n\\ 
\ea\right.
\eeq
has been used, and

\beq
\left\{\ba{ll}
{\bf T}_n &\!\!\equiv P_n(\cos\theta)\,{\bf t^{00}} \\
\vspace*{-3.5mm}\\
{\bf Z}_n &\!\!\equiv P_n^1(\cos\theta)\,{\bf t^{03}}  \\ 
\ea\right.
\quad
\left\{\ba{ll}
{\bf D}_n &\!\!\!\equiv P_n(\cos\theta)\,{\bf t^d}  \coma
{\bf H}^0_n\ \equiv P_n(\cos\theta)\,{\bf t^0}  \\
\vspace*{-3.5mm}\\
{\bf H}^1_n &\!\!\!\equiv P_n^1(\cos\theta)\,{\bf t^1}  \coma
{\bf H}^2_n\ \equiv P_n^2(\cos\theta)\,{\bf t^2}  \\
\ea\right.
\eeq
are the axially symmetric tensor  spherical harmonics. The constants $M_n^{(1)}$ and
$J_n^{(1)}$ re\-present the  linear intrinsic multipole moments of Geroch and Hansen
\cite{GH}, while the constants $A_n^{(1)}$ and $B_n^{(1)}$ are related to the stress
moments of the source in this linear approximation \cite{Elba}. The constants $A$ and $B$
could be eliminated by choosing an appropriate gauge, but they  are necessary for the
matching problem with the Lichnerowicz conditions \cite{Lichne}.

In the following we neglect the  multipoles $M_4^{(1)}$, $J_5^{(1)}$, $A_4^{(1)}$, $B_6^{(1)}$
and following. We also assume equatorial symmetry which eliminates the odd $(M, A, B)$ and
even $J$ multipole moments. This is coherent with the approximations done  in the last Section.
\subsection{Post linear or second order approximation (exterior solution)}

By constructing $\,N^{(2)}_{\a\be}\,$ and $\,H^{(2)}_\a$ from  (\ref{linearextsol})
and solving equations (\ref{plineareq}) for vacuum asuming the stationary and axial
symmetries, the following  expression for the metric to second order is obtained (the 
calculations has been performed using computer algebra) 

\beqa
{\bf g}^{[2]}_{\rm ext}
&\!\!\!\!=&\!\!\!\!\(-1+\frac{2M_0^{[2]}}{r}-\frac{2M_0^{(1)2}}{r^2}-
\frac{2M_0^{(1)}A_2^{(1)}}{3r^4}-
\frac{M_0^{(1)}B_2^{(1)}}{r^4}\){\bf T}_0 \nonumber\\
&\ \ +&\!\!\!\!\(\frac{2M_2^{[2]}}{r^3}-
\frac{4M_0^{(1)}M_2^{(1)}}{r^4}-\frac{4M_0^{(1)}A_2^{(1)}}{3r^4}\){\bf T}_2  +
\frac{2J_1^{[2]}}{r^2}{\bf Z}_1 +
\frac{2J_3^{[2]}}{r^4}{\bf Z}_3  \nonumber\\
&\ \
+&\!\!\!\!\(1+\frac{2M_0^{[2]}}{r}+\frac{4M_0^{(1)2}}{3r^2}
-\frac{2M_0^{(1)}A_2^{(1)}}{3r^4}-\frac{M_0^{(1)}B_2^{(1)}}{r^4}\){\bf D}_0 \nonumber\\ &\ \
+&\!\!\!\!\(\frac{2M_2^{[2]}}{r^3}+
\frac{64M_0^{(1)}M_2^{(1)}}{21r^4}-\frac{4M_0^{(1)}A_2^{(1)}}{3r^4}\){\bf D}_2 \nonumber\\
&\ \ +&\!\!\!\!\(\frac{A_2^{[2]}}{r^3}+
\frac{2M_0^{(1)}M_2^{(1)}}{7r^4}+\frac{2M_0^{(1)}A_2^{(1)}}{r^4}\){\bf E}_2 \nonumber\\
&\ \
+&\!\!\!\!\(\frac{B_2^{[2]}}{r^3}-\frac{M_0^{(1)2}}{3r^2}-
\frac{4M_0^{(1)}M_2^{(1)}}{21r^4}+\frac{2M_0^{(1)}B_2^{(1)}}{r^4}\){\bf F}_2\nonumber\\
&\ \
+&\!\!\!\!\(\frac{B_4^{[2]}}{r^5}-\frac{M_0^{(1)}M_2^{(1)}}{7r^4}\) {\bf F}_4
\label{plinearextsol} 
\eeqa
where we used the notation  $X^{[2]}\equiv X^{(1)} + X^{(2)}$ in order to take into
account the homogeneous part of the solution of (\ref{plineareq}), which is formally
identical to (\ref{linearextsol}). Moreover the terms involving $1/r^6$  or higher, as
well as the terms containing the products of $J$ moments by another moment, has been
omitted as they are not required for the final problem.
\section{Interior solution}
\subsection{The energy--momentum tensor}

Let us describe now the stellar model. The energy-momentum tensor describing the source is a
perfect fluid of constant density $\mu$ , with pressure $p$ depending  only on $r$ and 
$\th$, i.e. 

\beq
T_{\a\be} = \mu u_\a u_\be + p(g_{\a\be}+u_\a u_\be) \coma
\mu = {\rm const} \coma p(r,\theta)
\eeq
where $u_\a$ is the unit 4--velocity of the fluid. Further, we shall assume that no
convection takes place; so, $u_\a$ is a linear combination of the two Killing
vectors we are imposing

\beq
u^\a = \psi\(\xi^\a + \omega\z^\a\) 
 \coma \xi^\a = \de^\a_t \coma \z^\a =  \de^\a_\vf 
\eeq
Finally, we assume rigid rotation  (Born), that is, $\om$
 constant. Consequently, the normalization factor $\psi$ has the following
expression.

\beq
 \psi =
\frac1{\sqrt{-(g_{00}+2\om g_{03}r\sin\th+\om^2 g_{33}r^2\sin^2\th)}}
\eeq

On the other hand, the Euler equations can be written in this case as

\beq
\nabla_\a T^{\a\be} = 0 \quad \Ra \quad \frac{dp}{\mu+p} = d\log\psi
\eeq
which for constant density implies

\beq 
p = \mu\(\frac{\psi}{\psi_{_\S}}-1\)
\eeq
where $\psi_{_\S}$ is the value of $\psi$ on the surface of the fluid,
$p=0$. The equation defining this surface will therefore reads
\cite{Boyer}:  

\beq
\S:\ \psi(r,\theta) = {\rm const} \equiv \psi_{_\S} 
\eeq

Now, in order to handle the energy--momentum tensor in the postminkowskian
approximation context, it is necessary to asign an order of magnitude to the
constants appearing in this tensor: $\mu$, $\omega$ and $\psi_{_\S}$. First of all  we will
assume obviously that the {\it density is a first order\/} quantity (i.e. linear).
Furthemore, for reasons which will become clear in the last section  we will assume that the
{\it square of angular velocity is also of first order\/}. So, if we perform the following
substitutions  

\beq
\left\{\!\!\!\!\ba{ll}
& g_{00} \ra -1 + h^{(1)}_{00} \coma g_{03} \ra h^{(1)}_{03} \coma g_{33} 
= f_d + f_0 -f_2 \\
\vspace*{-2.5mm}\\
& f_d \ra 1 + h^{(1)}_d \coma f_0 \ra h^{(1)}_0\coma f_1 \ra h^{(1)}_1
\coma f_2 \ra h^{(1)}_2\\
\vspace*{-3mm}\\
&\psi_{_\S} \ra 1+ \psi^{(1)}_{_\S}
\ea\right.
\eeq
we finally get the  following tensor ${\bf{\cal T}}$ to second order (in the cylindrical
basis)

\beqa
{\cal T}^{[2]} &\!\!\!\!\simeq &\!\!\!\!\frac{\mu}2\[1 
+\frac12h^{(1)}_{00}-3\psi^{(1)}_{_\S}+
\frac72\om^2 r^2\sin^2\th \]{\bf t^{00}} 
-  \mu\, \omega r\sin\theta\, {\bf t^{03}} \nonumber\\
&+& \!\!\!\!\frac{\mu}2\[1 + h^{(1)}_d-\frac12h^{(1)}_{00}+\psi^{(1)}_{_\S}+
\frac16\om^2 r^2\sin^2\th \] {\bf t^d} \nonumber\\
&+&\!\!\!\!\frac{\mu}2\[h^{(1)}_0+  \frac13\om^2 r^2\sin^2\th\] {\bf
t^0}    +\frac{\mu}2 h^{(1)}_1{\bf t^1} 
+ \frac{\mu}2\[h^{(1)}_2-\omega^2 r^2\sin^2\theta\]{\bf t^2}
\eeqa
It should be remarked that terms containing $h^{(1)}_{03}$ has been neglected even though
they are postlinear in character. This is in agreement with the remark after
expression (\ref{plinearextsol}).
\subsection{Linear approximation (interior solution)}

Imposing regularity at the origin together with the usual symmetries, the general
solution of  equations (\ref{lineareq}) for the {\em linear part\/} of the 
energy--momentun tensor is 

\beqa
{\bf h}^{(1)}_{\rm int} &=& \!\!\!\sum_{n=0}^\infty
m_n^{(1)}\, r^n\,{\bf T}_n + \sum_{n=1}^\infty \,j_n^{(1)}\, r^n\,{\bf Z}_n 
 +\sum_{n=0}^\infty m_n^{(1)}\,r^n\,{\bf D}_n  \nonumber\\
&\ \ +&\sum_{n=0}^\infty a_n^{(1)}\, r^n\,{\bf\tilde E}_n   +\sum_{n=0}^\infty b_n^{(1)}\,
r^n\, {\bf\tilde F}_n + {\bf h}^{(1)}_{\rm p}
\eeqa
where:

\beq
\left\{\ba{ll}
{\bf\tilde E}_n &= {\bf E}_n +\dfrac23(2n+1) {\bf D}_n
-\dfrac23(2n+1){\bf H}^0_n \\
\vspace*{-3mm}\\
{\bf \tilde F}_n &= -(2n+1){\bf E}_n+ {\bf F}_n -\dfrac23n(2n+1){\bf D}_n
+\dfrac13(2n+1)(2n+3){\bf H}^0_n \\
\ea\right.
\eeq
and where ${\bf h}^{(1)}_{\rm p}$ is any particular solution. The constants $m_n^{(1)}$,
$j_n^{(1)}$, $a_n^{(1)}$ y $b_n^{(1)}$ have no specific meaning as in the exterior
case. However, since in the exterior we have neglected the multipoles
$M_4$, $J_5$, $A_4$, $B_6$ and following, the same will be done for the coefficients of
the analogous spherical harmonics  in the interior. So, we have  the following  linear
approximate interior solution 

\beqa
{\bf g}^{[1]}_{\rm int}
&&\hspace*{-5mm}=\[-1+m^{(1)}_0\]{\bf T}_0 +m^{(1)}_2\, r^2\,{\bf T}_2 +
j^{(1)}_1\,r\,{\bf Z}_1 + j^{(1)}_3\,r^3\,{\bf Z}_3\nonumber\\[1.5mm]
&&\hspace*{-3mm}+ \[1+m^{(1)}_0\]{\bf D}_0 + m^{(1)}_2\,r^2\,{\bf D}_2 
+a^{(1)}_0\,{\bf \tilde E}_0 +b^{(1)}_{0}\,{\bf \tilde F}_0 +\nonumber\\[1mm]
&&\hspace*{-3mm}+ a^{(1)}_2\,r^2\,{\bf\tilde E}_2 + b^{(1)}_2\,r^2\,{\bf\tilde F}_2 
\,\underbrace{-\frac43\pi\mu r^2\({\bf
T}_0+{\bf  D}_0\) -\frac85\pi\mu\,\om r^3\,{\bf Z}_1}_{\dps
{\bf h}^{(1)}_{\rm p}}
\eeqa
\subsection{Post linear approximation (interior solution)}

The postlinear approximation can be constructed in a similar way as for the exterior
case. This would leads to a similar expression to  (\ref{plinearextsol}), however it is
much longer and will not be written here. Nevertheless we shall write down below the final
expression after the corresponding  approximation has been made and the matching
with the exterior completed.
\section{The matching}
\subsection{Expansion of the constants and matching surface}

This subsection is the key to understand the analysis carried out in the present
paper. First we take an adimensional parameter  $\la$ which essentialy
reproduces  the post--minkowskian approximation  carried out above. Then
we use another adimensional parameter $\Om$ describing a slow rotation
approximation \cite{Hartle} \cite{Nikolaus} \cite{Thorne} whose first term has spherical
symmetry (``post--spherical approximation"). Specifically  the parame\-ters we have used
are: 

A) $\la$  is the ratio between $m$, a constant with units of mass, later
 identified as the ``newtonian mass" of the source, and a length $r_0$ measuring the
radius of the star in  absence of rotation (i.e. spherical symmetry). 

B) $\Om$ is the ratio between the centrifugal and gravitational forces at the surface of the
star. It is therefore proportional to the rotational velocity of the source and measures its
deviation from spherical symmetry. This parameter is used to perform, at each order of
$\la$, a second approximation (slow rotation).

\beq
\left\{\!\!\!\!\!\!\ba{ll}
&\la = \dfrac{m}{r_0}\,:\ {\rm Post\!-\!minkowskian\ approximation} \qquad \(m=
\dfrac43\pi\,r_0^3\,\mu\)\\
\vspace*{-2mm}\\
&\om r_0 = \la^{1/2} \Om \coma \Om\,:\ {\rm Slow\ rotation\ approximation}\\
\ea\right.
\eeq
Notice that the exponents of $\la$ for the ${\bf Z}_n$ coefficients will be
half-integers. Nevertheless, with respect to the post--minkowskian approximation the order
of magnitude of the corresponding terms will equal the integer part of those
exponents. The final approximation will consist on imposing  $\,\la^{5/2}
\ra 0\,$ and $\,\Om^4 \ra 0\,$.

The fundamental hypothesis of this work is demonstrated in the following
expansions for the ``exact" constants $M_n$, $J_n$, $A_n$ and $B_n$ (the formal meaning
of  the term  ``exact" in our case is  $X_n =\lim_{r \to \infty}X_n^{[r]}$). In general,
these are  expansions in powers of $\la$, each coefficient of which is, in turn, expanded
in powers of $\Om$. This hypothesis is based on the newtonian analysis of a
selfgravitating rotating object, which has, as is well--known, the Maclaurin spheroids as
possible solutions \cite{Luminet}.

\beqa
&&\dfrac{M_0}{r_0} = \la\[1 +M_0^{(1,2)}\Om^2 +O(\Om^4)\] +\la^2\[M_0^{(2,0)}
+M_0^{(2,2)}\Om^2+O(\Om^4)\] +O(\la^3) \nonumber\\[2mm]
&&\dfrac{M_2}{r_0^3} = \la\Om^2\[M_2^{(1,2)} +O(\Om^2)\] + \la^2\Om^2\[M_2^{(2,2)}
+O(\Om^2)\] +O(\la^3)  \nonumber\\[2mm]
&&\dfrac{J_1}{r_0^2} = \la^{3/2}\Om\[J_1^{(1,1)} +J_1^{(1,3)}\Om^2+O(\Om^4)\] 
+O(\la^{5/2}) \nonumber\\[2mm]
&&\dfrac{J_3}{r_0^4} = \la^{3/2}\Om^3\[J_3^{(1,3)} +O(\Om^2)\]  +O(\la^{5/2})
\eeqa
\beqa
&&\dfrac{A_2}{r_0^3} = \la\Om^2\[A_2^{(1,2)} +O(\Om^2)\]  +\la^2\Om^2\[A_2^{(2,2)}
+O(\Om^2)\] +O(\la^3)  \nonumber\\[2mm]
&&\dfrac{B_2}{r_0^3} = \la\[B_2^{(1,0)} +B_2^{(1,2)}\Om^2 +O(\Om^4)\] +
\la^2\[B_2^{(2,0)} +B_2^{(2,2)}\Om^2+O(\Om^4)\] +O(\la^3) \nonumber\\[2mm]
&&\dfrac{B_4}{r_0^5} = \la\Om^2\[B_4^{(1,2)} +O(\Om^2)\]  +\la^2\Om^2\[B_4^{(2,2)}
+O(\Om^2)\] +O(\la^3) 
\eeqa

For  $M_0$ and $B_2$, the first terms are independent of
$\Om$ because they already appear in spherical symmetry \cite{Elba}
\cite{Quan}. For  $J_1$ (angular momentum) we can assume that the
lowest term is linear in $\Om$, which corresponds to a sphere put into rotation.
However, $M_2$ and $J_3$ represent true deformations of sphericity due to rotation;
therefore, we assume that the expansions start as indicated. Finally, for
$A_2$ and $B_4$, we assume a similar  structure to $M_2$; this is based on the fact
that  these constants are associated to the spherical harmonics of the same order as
the quadrupole moment.

Regarding the constants $m$, $j$, $a$ and $b$  in the interior solution, we
take similar expansions as above. For shortness they will not be included here.

In agreement with the ``post--spherical approximation" we make the suplementary hypothesis
that the equation determining the surface $\S$ of the star has the  structure:

\beq
r = r_0 + r_2\Om^2 P_2(\cos\th) + O(\Om^4)
\eeq
so that, for vanishing rotation, it becomes the equation of a sphere of radius $\,r_0$. 
In the following subsections we analize the matching problem. Let us advance here the
solution we find for this surface, which reads:

\beq
\S:\ \eta = 1 +\(-\frac56 +\frac{5\la}7\)\Om^2\,P_2(\cos\th)+ O(\Om^4)\coma 
\eta \equiv \frac{r}{r_0}
\eeq

\subsection{Results of the matching (exterior solution)}

Imposing the condition that the metric (interior and exterior) is
continuous with continuous first derivatives on the surface of the fluid,
the following expansions for the constants in the exterior solution are
found:

\beq
\left\{\!\!\ba{ll}
\dfrac{M_0}{r_0} &\!\!\!=\, \la + \la^2\(3 +\dfrac25\Om^2\) +O(\la^3,\Om^4)\\[4mm]
\dfrac{M_2}{r_0^3} &\!\!\!=\, -\dfrac12\la\Om^2 -\dfrac{69}{70} \la^2\Om^2
+O(\la^3,\Om^4)\\
\ea\right.
\left\{\!\!\ba{ll}
\dfrac{J_1}{r_0^2} &\!\!\!=\, \la^{3/2}\Om\(\dfrac25+\dfrac13\Om^2\)  +O(\la^{5/2},\Om^5)
\\[4mm]
\dfrac{J_3}{r_0^4} &\!\!\!=\,-\dfrac17\la^{3/2}\Om^3  +O(\la^{5/2},\Om^5) \\
\ea\right.
\eeq

\beq
\left\{\!\!\ba{ll}
A_2 &\!\!\!=\,  O(\la^3,\Om^4) \\[2mm]
\dfrac{B_2}{r_0^3} &\!\!=\, \la^2\(\dfrac8{35} +\dfrac4{105}\Om^2\) +O(\la^3,\Om^4)
\\[4mm]
\dfrac{B_4}{r_0^5} &\!\!\!=\, -\dfrac4{63}\la^2\Om^2 +O(\la^3,\Om^4) \\
\ea\right.
\eeq
In the $M_0$ expansion, the different contributions (newtonian mass, binding energy,
rotational energy) to the total mass of the body can be observed.

Plugging this expressions  into (\ref{plinearextsol}) the following metric tensor of the
exterior region is obtained

\beqa
{\bf g}_{\rm ext}^{[2]} &\!\!\!\!\!\!\!= &\!\!\!\!\!\!\!\[-1+\frac{2\la}\eta
+2\la^2\(3 +\frac{2\Om^2}{5}\)\frac1\eta-\frac{2\la^2}{\eta^2}\]\!{\bf T}_0
+\(-\frac{\la}{\eta^3} -\frac{69\la^2}{35\eta^3} +\frac{2\la^2}{\eta^4}\)
\!\Om^2\,{\bf T}_2  \nonumber\\[2mm]
&+&\!\!\!\!2\la^{3/2}\Om\(\frac2{5} +\frac{\Om^2}{3}\)\!\frac{1}{\eta^2}\,{\bf Z}_1
-\frac{2\la^{3/2}\Om^3}{7\eta^4}\, {\bf Z}_3 \nonumber\\[2mm]
&+&\!\!\!\!\[1+\frac{2\la}{\eta}+2\la^2\(3
+\frac{2\Om^2}{5}\)\frac1\eta
+\frac{4\la^2}{3\eta^2}\]\!{\bf D}_0 
 +\(-\frac{\la}{\eta^3} -\frac{69\la^2}{35\eta^3}
-\frac{32\la^2}{21\eta^4}\)
\!\Om^2\,{\bf D}_2    \nonumber\\[2mm]
&-&\!\!\!\!\frac{\la^2\Om^2}{7\eta^4}\,{\bf E}_2+\la^2\[-\frac{1}{3\eta^2}
+\frac4{35}\(2 +\frac{\Om^2}{3}\)\frac1{\eta^3}
+\frac{2\Om^2}{21\eta^4}\]\!{\bf F}_2 \nonumber\\[2mm]
&+&\!\!\!\!\la^2\Om^2\(\frac{1}{14\eta^4}-
\frac{4}{63\eta^5}\){\bf F}_4
\eeqa
We should remark that, as mentioned in the abstract, the Kerr metric does not belong
to this approximate solution. Kerr's metric can be obtained from (\ref{plinearextsol}) by
making the substitutions $M_0 = m$, $M_2=-ma^2$, $J_1=ma$ and $J_3=-\frac13 ma^3$ (here
$m$ and
$a$ stand for the standard parameters of the Kerr metric). It can be easily checked
that these values  can not be obtained by any choice of the free
parameters
$\la$, $\Om$ and $r_0$. One might think that this need not to be true for another fluid,
for instance a polytrope or differential rotation. 

\subsection{Results of the matching (interior solution)}

Similarly, the matching conditions produce the following expressions for the
constants of the interior solution

\beq
\!\!\!\left\{\!\!\!\!\!\!\!\!\ba{ll}
&m_0 = 3\la + \dfrac14\la^2(21 +2\Om^2) +O(\la^3,\Om^4)\\[2mm]
&m_2r_0^2 = -\la\Om^2 -\dfrac{73}{70}\la^2\Om^2 +O(\la^3,\Om^4)\\
\ea\right. 
\left\{\!\!\!\!\!\!\!\!\ba{ll}
&j_1r_0 = \la^{3/2}\Om\(2+\dfrac23\Om^2\)  +O(\la^{5/2},\Om^5) \\[2mm]
&j_3r_0^3 = -\dfrac27\la^{3/2}\Om^3  +O(\la^{5/2},\Om^5) \\
\ea\right.
\eeq

\beq
\left\{\!\!\!\!\!\!\!\ba{ll}
&a_0 =  \dfrac13\la^2(21 +2\Om^2) + O(\la^3,\Om^4)\\[2mm]
&a_2r_0^2 =  -\dfrac{43}{15}\la^2 \Om^2 + O(\la^3,\Om^4) \\
\ea\right.
\left\{\!\!\!\!\!\!\!\ba{ll}
&b_0 = O(\la^3,\Om^4) \\[2mm]
&b_2r_0^2 = -\dfrac{86}{105}\la^2\Om^2 +O(\la^3,\Om^4) \\ 
\ea\right.
\eeq
so that the metric in the interior region once matched to the exterior reads:

\beqa
{\bf g}_{\rm int}^{[2]} &\!\!\!\!\!=& \!\!\!\!\!\!\[-1+3\la-\la\eta^2
+\frac12\la^2\(\frac{21}2 +\Om^2\) +\la^2\(-\frac32+\Om^2\)\eta^2
+\frac12\la^2\(\frac12 -\frac75\Om^2\)\eta^4\]\!{\bf T}_0 \nonumber\\[2mm]
&\!\!\!\!\!\!\!\!\!\!\!\!+&\!\!\!\!\(-\la -\frac{73}{70}\la^2
+\frac{15}{14}\la^2\eta^2\)\Om^2\!\eta^2\,{\bf T}_2  +\la^{3/2}\[2 +\frac23\Om^2
-\frac65\eta^2\]\Om\,\eta {\bf Z}_1 -\frac27\la^{3/2}\Om^3\eta^3\, {\bf Z}_3 
\nonumber\\[2mm] 
&\!\!\!\!\!\!\!\!\!\!\!\!+&\!\!\!\!\[1+3\la-\la\eta^2 +\frac12\la^2\(\frac{49}2
+\frac73\Om^2\) -\la^2\(\frac{11}2+\frac13\Om^2\)\eta^2 +\frac16\la^2\(\frac7{2}
 -\frac1{5}\Om^2\)\eta^4\]\!{\bf D}_0  \nonumber\\[2mm]
&\!\!\!\!\!\!\!\!\!\!\!\!+&\!\!\!\!\(-\la\eta^2 -\frac{1079}{210}\la^2\eta^2
+\frac{23}{14}\la^2\eta^4\)\!\Om^2\,{\bf D}_2
+\frac17\la^2\Om^2\(-4\eta^2 +3\eta^4\){\bf E}_2  \nonumber\\[2mm]
&\!\!\!\!\!\!\!\!\!\!\!\!+&\!\!\!\!\la^2\[\frac15\(-1+\frac{18}{7}\Om^2\)\eta^2 +
\frac2{21}\(1-4\Om^2\)\eta^4\]\!{\bf F}_2
+\frac1{126}\la^2\Om^2\eta^4\,{\bf F}_4
\eeqa

\vfill\eject
\noindent On the other hand the expression for the pressure is the following:

\beq
p = \dfrac{\la^2}{8\pi r_0^2}\Big[(3-2\Om^2)(1-\eta^2)-5\Om^2\eta^2 P_2(\cos\th)\Big] + O(\la^3,\Om^4)
\eeq

\section{Acknowledgments}

This work has been supported by a Research Grant from the Spanish ``Ministerio de
Ciencia y Tecnolog\'\i{}a", number BFM2000--1322. The authors thank J.M.
Aguirregabir\'\i{}a and M. Mars for helpful comments and discussions.

\end{document}